\documentclass[12pt]{article}

\usepackage{scicite}

\usepackage{times}

\usepackage{graphicx}
\usepackage{xcolor}

\newenvironment{sciabstract}{%
\begin{quote} \bf}
{\end{quote}}

\newcounter{lastnote}

\title{Magnetic measurements on micron-size samples under high pressure using designed NV centers}

\author
{Margarita Lesik$^{1}$\footnote{These authors contributed equally to this work.}, Thomas Plisson$^{2,\ast\dagger}$, Lo\"\i c Toraille$^{1,\ast}$, 
Justine Renaud$^{3}$,  \\
Florent Occelli$^{2}$, Martin Schmidt$^{1}$, 
Olivier Salord$^{3}$, Anne Delobbe$^{3}$, \\ Thierry Debuisschert$^{4}$,
  Lo\"\i c Rondin$^{1}$, Paul Loubeyre$^{2}$, 
and Jean-Fran\c cois Roch$^{1,\dagger}$\\
\normalsize{$^{1}$Laboratoire Aim\'e Cotton, CNRS, Univ. Paris-Sud, ENS  Cachan,\linebreak}\\
\normalsize{$^{}$Universit\'e Paris-Saclay, 91405, Orsay Cedex,  France}\\
\normalsize{$^{2}$CEA, DAM, DIF, 91297 Arpajon, France}\\
\normalsize{$^{3}$Orsay Physics S. A., 95 avenue des Monts Aur\'eliens, 13710 Fuveau, France}\\
\normalsize{$^{4}$Thales Research \& Technology, 1, avenue Augustin Fresnel, 91767 Palaiseau, France}
\\
\normalsize{$^\dagger$To whom correspondence should be addressed;} \\
\normalsize{E-mail:  thomas.plisson@cea.fr, jean-francois.roch@ens-paris-saclay.fr.}
}

\date{}

\begin{document} 


\baselineskip24pt


\maketitle 

% Abstract should be 125 words or less

 \begin{sciabstract}
 Pressure is a unique tool to tune the interplay between structural, electronic and magnetic interactions. It leads to remarkable properties of materials such as recent temperature records in superconductivity. Advanced magnetic measurements under very high pressure in the Diamond Anvil Cell (DAC) use synchrotron approaches but these are lacking a formal link to the macroscopic magnetic properties. We report an alternative method consisting in optical magnetometry based on nitrogen-vacancy (NV) centers created at the surface of a diamond anvil. We illustrate the method by two measurements realized at  room and low temperature respectively: the pressure evolution of the magnetization of an iron bead up to 30~GPa showing the iron ferromagnetic collapse and the detection of the superconducting transition of MgB$_2$ at 7~GPa.
\end{sciabstract}

\pagebreak
   
Compression of a solid directly changes its electron density inducing a large diversity of magnetic phenomena  such as quantum critical points or high-spin low-spin transitions  \cite{Mao2018}.  Pressure is also a relevant  physical parameter for tuning superconductivity in a wide range of systems, such as cuprates, transition-metal dichalcogenides, iron pnictides, heavy fermions or topological superconductors. Recently a novel class of high temperature superconductors has been discovered in H-rich hydrides at high pressure, highlighted by the reported critical temperature of 203~K in H$_3$S at 150~GPa \cite{Drozdov2015}. 
Systematic exploration of these materials requires sensitive magnetic characterization that can be routinely operated in the 100 GPa range \cite{Gorkov2018}.

Great efforts have already been made to adapt magnetic detection methods  to the DAC specificities \cite{Eremets1996}. The macroscopic magnetic susceptibility of  the compressed sample can be measured using inductively coupled coils. Two strategies have been followed to circumvent the poorly scaling filling factor under pressure, detrimental for the detection sensitivity. One is to insert the detection coil inside the sample chamber \cite{Jackson2003} and the other to miniaturize the whole DAC to integrate it in a SQUID  in order to benefit from the  intrinsic high  sensitivity of SQUID measurements \cite{Marizy2017}. Synchrotron based methods, such as x ray magnetic circular dichroism (XMCD),  x ray emission spectroscopy  (XES), and nuclear resonant forward scattering (NRFS) are also widely used due to the development of  focused and high brightness synchrotron x ray beams. By addressing atomic or nuclear resonance lines, these methods are element specific  and resolve local magnetism due to a given electronic order \cite{Torchio2011}. However, quantitative analysis exploiting these techniques may be challenging \cite{Mathon2004,Rueff2010}. Furthermore, synchrotron-based techniques, mainly probing nuclear or electronic transitions, are not directly sensitive to magnetic phenomena such as the Meissner effect which is the signature of superconductivity. 

We  report an optical  magnetometry method  based on  NV  color centers  used as in situ quantum sensors (Fig.~1a). The main advantages are the easiness of the sample preparation, the  table-top instrumentation, the mapping of the stray magnetic field with micrometer spatial resolution  and the absence of any sensitivity decrease with the sample size, down to the micrometer scale. Being non invasive, the scheme of this  magnetic detection can be easily associated to complementary structural determination by x ray diffraction. 
The method is based on Optically Detected Magnetic Resonance (ODMR) which exploits the triplet  fine structure of the negatively charged NV$^-$ center ground state (Fig.~1b). ODMR relies on the difference in luminescence intensity when exciting the NV$^-$  center from $m_S=0$ (high intensity) or from $m_S=\pm 1$ (low intensity) spin states \cite{Doherty2013review}. This property is used to perform a spectroscopy of the $m_S=0 \to \pm 1$ transition excited by a microwave (MW) signal which frequency  is  scanned across the resonance. Under a continuous MW excitation in the absence of external perturbation, the spin dependent luminescence  recorded as a function of the MW frequency  exhibits  a resonance peak at 2.87~GHz, as shown in Fig.~1c. An external magnetic field applied on the NV center then splits this  resonance, leading to the direct measurement of the magnetic field amplitude from the ODMR spectrum \cite{Taylor2008}. The  influence  of strain  on this transition was investigated  up to a  pressure of about 60~GPa by Doherty \textit{et al}  and the application to magnetic detection in a DAC  was envisioned as the main goal \cite{Doherty2014}. Here, we fabricate  the NV centers  in the diamond anvil and we elaborate an analysis that unlocks magnetometry.  We illustrate its efficiency on two testbed examples: a quantitative measurement of the magnetization of iron up to 30~GPa under ambient temperature around the $\alpha \to \epsilon$  phase transition  \cite{Bassett1987} and the detection of the superconductivity of MgB$_2$ at 7~GPa and 30~K \cite{Buzea2001}	.

A focused ion beam extracted from a nitrogen plasma is used to create  the NV centers at the culet of an ultra-pure synthetic   (IIas) diamond anvil \cite{Lesik2013}, with   a layer of about $10^4$ defects per $\mu$m$^2$ at a depth of   20 nm below the surface of the anvil culet (see Supporting Information). As shown in Fig.~1a, the optical excitation of these shallow   NV centers using a laser of 532 nm wavelength  and the detection of their luminescence are performed  through the anvil. 
We then implemented the wide-field ODMR scheme of \cite{Steinert2010} to record the  map of the NV centers spin-state dependent luminescence  as a function of microwave frequency. The implanted diamond anvil is  mounted on a membrane DAC and a rhenium gasket laterally confines the sample. A single turn coil is positioned on the gasket as a MW antenna for the ODMR (Fig.~1a). A ruby crystal is used as a pressure gauge. Microscopic samples of iron or MgB$_2$   were positioned in the sample chamber, directly in contact with the implanted anvil (see Fig. 2), and are embedded in a pressure transmitting medium consisting respectively of nitrogen and argon.

 In the case of iron, various samples have been loaded to test geometric and size effects. The detailed analysis focuses on one of the iron beads  (see Supporting Information for the signals associated to the other samples).  A   magnetic field $B_0 \approx$~11~mT, created by a permanent magnet, is applied to magnetize the iron samples and to split and resolve the resonances associated to the four  orientations of the NV centers existing in a $\lbrack100\rbrack$-oriented diamond \cite{Toraille2018}. 

The energy levels of a given NV center  are modified by  the  magnetic field    and by the  strain field existing in the anvil. The combination of these two  perturbations   results in both a  shift and a splitting  of the MW resonance frequency \cite{Teissier2014}.  The combined effect of strain and magnetic field has been described by Barson et al \cite{Barson2017}. As sketched in Fig.~1b, the hydrostatic component of the strain shifts the resonance frequency whereas its non-hydrostatic component and the magnetic field both split the resonance around its center frequency. Extracting the magnetic field created by the iron bead magnetization from these resonances recorded at high pressure requires to take into account the competing effects of the magnetic field and of the strain field   which add quadratically \cite{Barson2017}. 
A typical spectrum obtained in the experiment is shown in Fig. 1c. At first order, the effect of the magnetic field  is proportional to its  longitudinal component    on the N-V axis. It then leads to a spectrum consisting of a   set of four double resonances corresponding to the four NV  orientations.  A map of the measured raw frequency splittings for each family of NV centers is plotted in Fig.~2 at a pressure of 24~GPa. The  splitting induced by   the iron bead magnetization is of the order of a few MHz, which can be resolved over the   splitting   induced by  $B_0$. Even prior to any   data analysis,  such images can be recorded live during the experiment providing   direct  evidence of    pressure induced modifications of the magnetic properties.

After  ascribing each pair of resonances to a given $\lbrack 111 \rbrack$  axis of the diamond anvil, the stray magnetic field created by the iron bead can be quantitatively extracted from the correlated informations that are embedded in the spectrum shown in Fig.~1c \cite{Chipaux2015}. Beforehand, the strain component is extracted from the ODMR signal to sort out the stray magnetic field of the iron bead by using a reference area of the image selected far from the bead. It is also observed that right below the bead, the NV centers undergo a strong transverse magnetic field which mixes the  $m_S=\pm 1$ states and induces a strong decrease of the ODMR contrast~\cite{Tetienne2012}. This quenching leads   to a non-relevant background signal without a direct link to the iron bead magnetization. As shown in Fig. 3b, the area below the center of the bead is excluded by applying a mask to the data according to a threshold set on the contrast (see Supporting Information for the data processing).   At high pressure  where the bead magnetization decreases, the masked area decreases accordingly.

The magnetization $M$ of the iron bead is then determined by fitting the relevant part of the magnetic field map with a simple magnetic dipole model (see Supporting Information).  At low pressure, we obtain $M \approx 8 \pm 1 $ kA$\cdot$m$^{-1}$. The evolution of the magnetization with pressure is shown in Fig.~3a. The magnetic field of the iron bead decreases as the pressure increases from  15~GPa  up to about  30~GPa (Fig.~3b), above which no significant value of the magnetization can be  inferred. This result  demonstrates the sensitivity of the   detection scheme and is consistent with previous XES and SQUID measurements that reported magnetic signals well above the $\alpha-\epsilon$ transition threshold \cite{Monza2011,Wei2017}.  Whether a remaining of the $\alpha$ phase is responsible for this magnetic signature up to 30~GPa could be investigated by implementing  NV-based magnetometry on an x ray diffraction beamline, combining  structural and magnetic measurements on the same sample at each pressure. Finally, the pressure in the DAC was released; we then  observed the reappearing of the sample magnetization with the expected hysteresis behavior  related to the hysteresis of the structural transition  \cite{Bassett1987}. 

This technique can be straightforwardly implemented   at low temperatures in order to observe the Meissner effect associated to a superconductor. As a testbed system, we chose a sample of MgB$_2$   that was confined  in the DAC at 7~GPa and first cooled in a cryostat in the absence of external magnetic field. At a temperature of about 18 K, an external magnetic field $\approx 1.8$~mT  is applied. The direction of this magnetic field is chosen parallel to the $\lbrack 100 \rbrack$ diamond axis so that the four NV orientations in the crystal have  identical responses. As shown in Fig.~4a, the comparison between the frequency shift in the   electron spin resonance (ESR) of the NV centers located above the sample and the homogeneous background  gives a direct image of the Meissner effect then providing a direct proof of  MgB$_2$ superconductivity (Fig.~4a). 
 Under heating of the DAC, vanishing of the superconductivity is detected above 30~K  (Fig.~4b). This behavior agrees with  the reported pressure evolution of the MgB$_2$ critical temperature  \cite{Buzea2001}. 

The next step is to investigate how this direct detection of the Meissner effect can be implemented at a pressure range above 100 GPa. This may require adapting the excitation and readout wavelengths   of the NV center in order to compensate for the pressure induced energy shifts    of the electronic levels   \cite{Doherty2014}. NV engineering based on controlled nitrogen doping during the plasma-assisted growth of a  diamond layer  \cite{Lesik2016} or using laser writing \cite{Chen2017} can be used to bury a thin sheet of NV centers at a depth where the influence of strain in the anvil could be less detrimental. A foreseen major application  is the investigation of high-temperature superconductivity  in the various super-hydride compounds that can be directly synthesized at high pressure, such as H$_3$S~\cite{Drozdov2015}, UH$_7$~\cite{Kruglov2018}, LaH$_{10}$~\cite{Geballe2018}  and FeH$_5$~\cite{Pepin2017}. Finally, it could enable the observation of the predicted magnetic properties of metallic hydrogen which is within experimental reach \cite{Dias2017_Comments} and for which various challenging experimental probes have already been proposed \cite{Babaev2005,Carbotte2018}.

\clearpage

\begin{figure}  [h!]
 \centerline{\includegraphics[width=13.5cm]{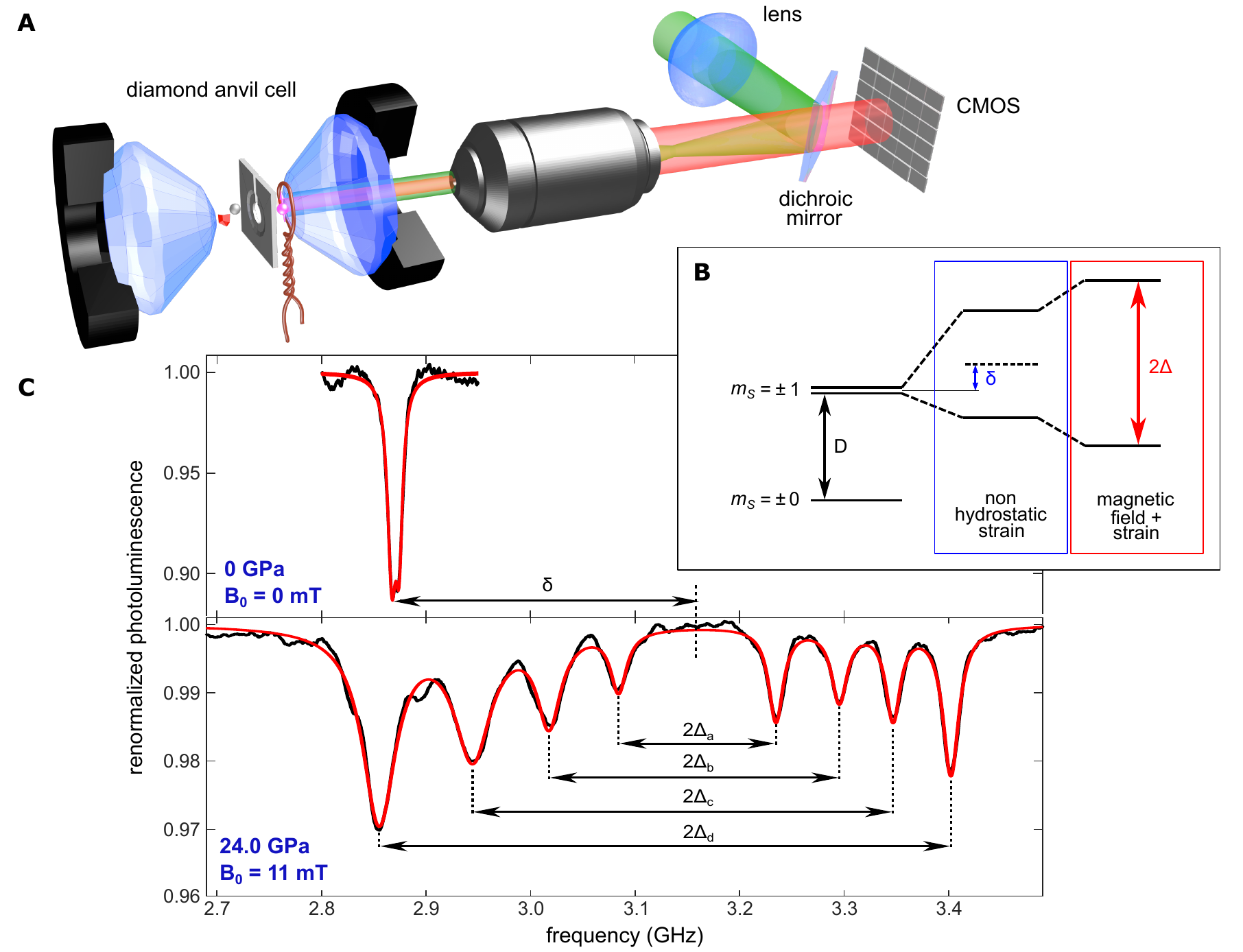}}
\end{figure}
\noindent {\bf Fig. 1. Implementation of NV magnetometry in a DAC.}  (\textbf{A}) Scheme of the DAC setup. A 250~$\mu$m-wide disk of NV centers is implanted in the 300~$\mu$m-wide culet of one of the two anvils. A rhenium gasket encloses the sample and a ruby pressure gauge. A single-loop wire is placed on the gasket to generate the microwave excitation. A  laser with 532 nm wavelength is used to excite the luminescence of the NV centers. The electron spin resonance   is detected through its effect on the luminescence by using a CMOS camera. (\textbf{B}) Electronic structure of the NV center ground state with the modification of the energy levels 
induced by the strain in the anvil and the   magnetic field.   (\textbf{C}) Typical resonance spectra. In the absence of any perturbation the spectrum consists of a single resonance at  D=2.87~GHz. Hydrostatic compression  shifts this resonance by $\delta$ while   non-hydrostatic strain splits the resonance in two components that are also affected by the  applied magnetic field with a total frequency splitting $2 \Delta$.  The projections of the magnetic field on the  four NV orientations  in the crystal lead to eight resonance peaks.   

\clearpage

\begin{figure}  [h!]
 \centerline{\includegraphics[width=9cm]{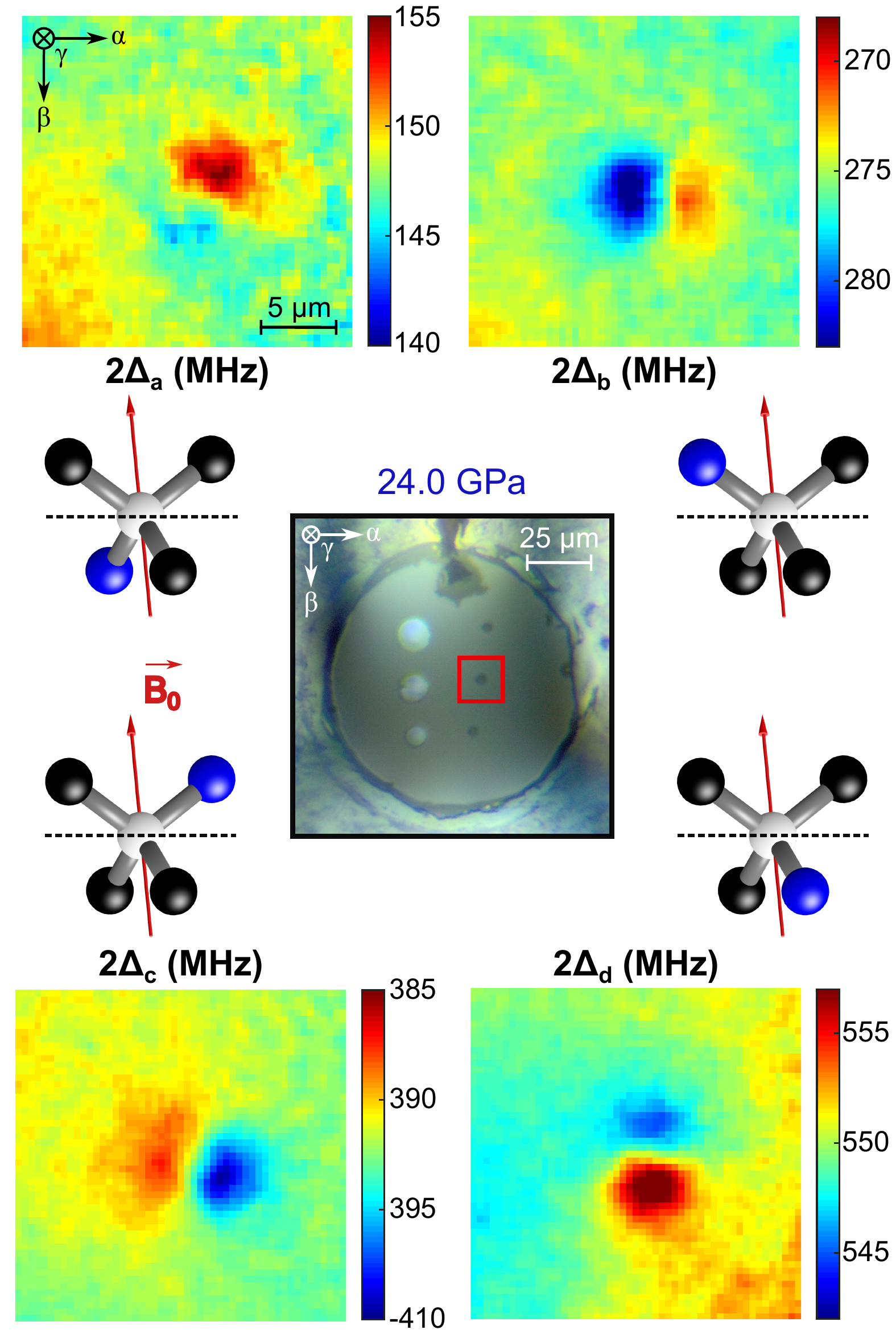}}
\end{figure}
\noindent {\bf Fig. 2. Frequency splittings associated to  the magnetization of the iron bead at 24~GPa for the four NV orientations.} The data are shown for   the  bead indicated  by the red square in the center image showing the iron samples inside the  gasket. The splittings combine the effect of non hydrostatic strain in the anvil, of the applied magnetic field  and of the stray magnetic field created by the bead magnetization. $(\alpha, \beta, \gamma)$ are  reference axes linked to the anvil used to identify the four NV orientations. The dotted lines indicate the orientation of the anvil surface.

\clearpage

\begin{figure}  [h!]
 \centerline{\includegraphics[width=9cm]{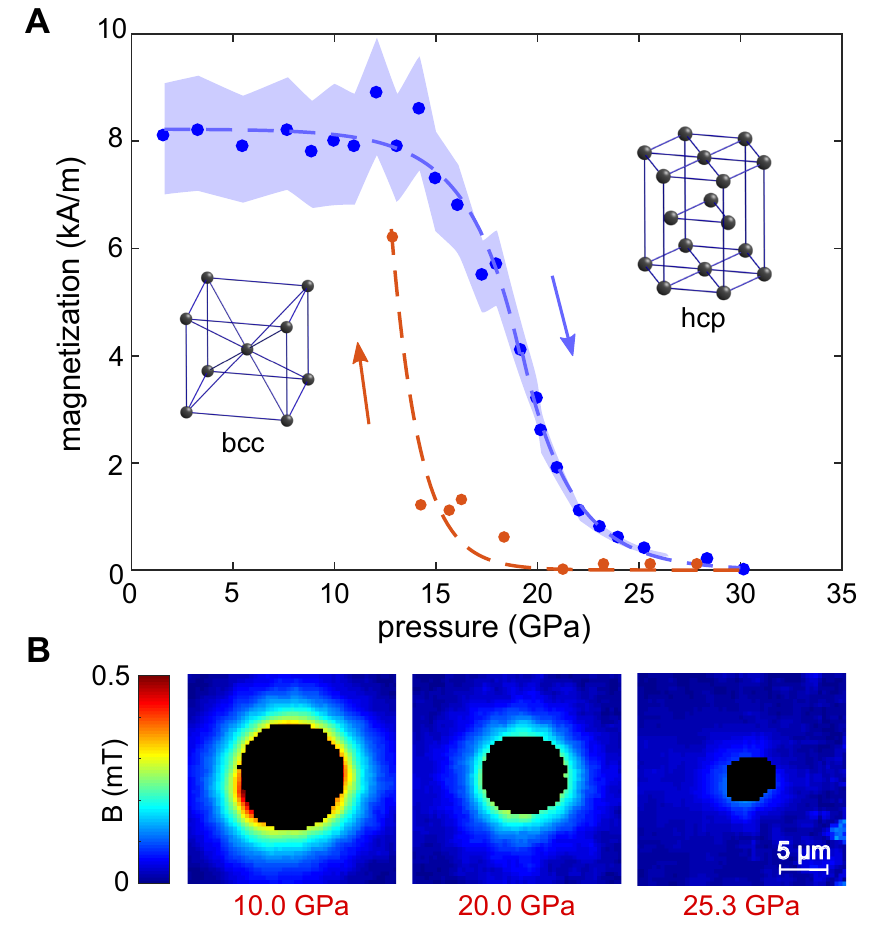}}
\end{figure}
\noindent {\bf Fig. 3. Observation of the $\alpha-\epsilon$   transition of iron associated to (bcc) and (hcp) crystal structures.} (\textbf{A}) Evolution of the  bead magnetization inferred at each pressure. The data taken during the pressure increase is shown by blue dots while the data taken during the pressure release is shown by red dots. The shaded area shows the uncertainty interval on the magnetization value during the pressure increase. The dotted lines are simple guides to the eye. (\textbf{B}) Evolution of the  amplitude of the magnetic field  created by the iron bead. The mask shown in black is associated to the criteria set on the ODMR contrast.  

  \clearpage

\begin{figure}  [h!]
 \centerline{\includegraphics[width=18.4cm]{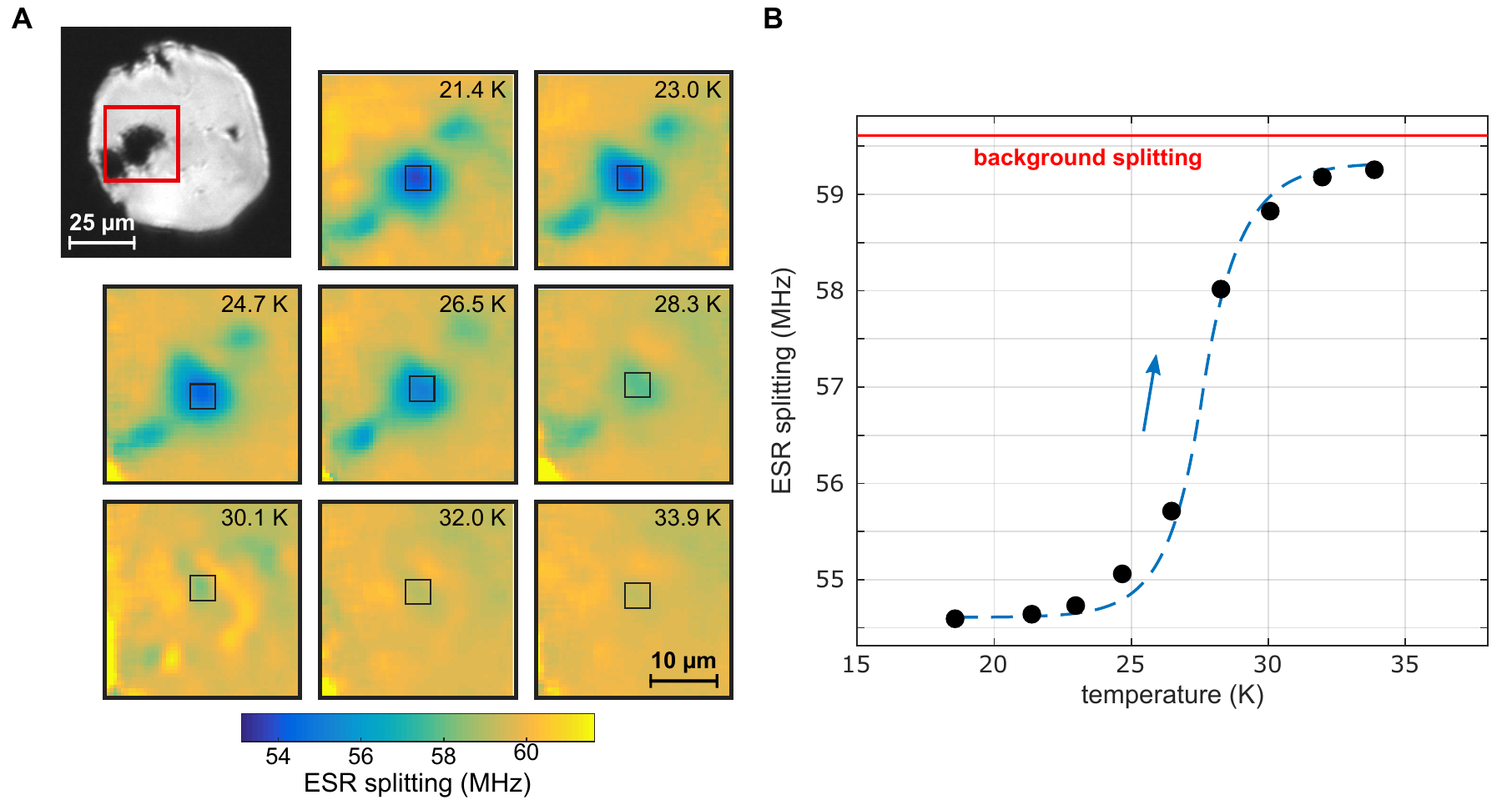}}
\end{figure}
\noindent {\bf Fig. 4. Meissner effect associated to the superconductivity of MgB$_2$ at 7~GPa.} (\textbf{A}) Maps of the ESR frequency splitting above the MgB$_2$ sample recorded for an increasing temperature. A control magnetic field $B_0\approx 1.8 \, {\rm mT}$ is applied and induces   a background ESR splitting which combines the influence of $B_0$ with the strain in the anvil. Below 30~K, the exclusion of the magnetic field is observed above the MgB$_2$ sample   due to the Meissner effect. This effect disappears   above 30~K leading to an homogeneous distribution of the ESR splitting. Inset: optical image of the sample. The red square indicates where the ESR splitting is mapped.   (\textbf{B}) Evolution of the ESR splitting  above the  sample when the temperature is increased. The data points are averaged on the black squares of (\textbf{A}). The dotted line is a  guide to the eye. 

 \clearpage

\end{document}